# The use of restricted mean time lost under competing risks data[§]

**Jingjing Lyu[1], Yawen Hou[2], Zheng Chen[1,*]**

[1]Department of Biostatistics, School of Public Health, Southern Medical University. No. 1023, South Shatai Road, Baiyun District, Guangzhou 510515, China.

[2]Department of Statistics, College of Economics, Jinan University, Guangzhou, China.

∗ Corresponding author

**Abstract**

**Background:** Under competing risks, the commonly used sub-distribution hazard ratio (SHR) is not easy to interpret clinically and is valid only under the proportional sub-distribution hazard (SDH) assumption. This paper introduces an alternative statistical measure: the restricted mean time lost (RMTL).

**Methods:** First, the definition and estimation methods of the measures are introduced. Second, based on the differences in RMTLs, a basic difference test (Diff) and a supremum difference test (sDiff) are constructed. Then, the corresponding sample size estimation method is proposed. The statistical properties of the methods and the estimated sample size are evaluated using Monte Carlo simulations, and these methods are also applied to two real examples.

**Results:** The simulation results show that sDiff performs well and has relatively high test efficiency in most situations. Regarding sample size calculation, sDiff exhibits good performance in various situations. The methods are illustrated using two examples.

**Conclusions:** RMTL can meaningfully summarize treatment effects for clinical decision making, which can then be reported with the SDH ratio for competing risks data. The proposed sDiff test and the two calculated sample size formulas have wide applicability and can be considered in real data analysis and trial design.

**KEY WORDS**: Competing risks; Hypothesis tests; Proportional sub-distribution hazard assumption; Restricted mean time lost

# 1. Background

Competing risks arise frequently in many applications in medical studies. In a competing risks setting, patients may fail due to multiple causes. The most commonly researched endpoint is recorded as the event of interest; other endpoints, whose occurrence may preclude the occurrence of the event of interest, are recorded as competing events [1]. When competing risks exist, the Kaplan-Meier estimation tends to overestimate the cumulative incidence function, which may cause large errors and lead to incorrect conclusions [2, 3]. The commonly used measures in the present competing risks data analysis are the cumulative incidence function (CIF), sub-distribution hazard (SDH), and cause-specific hazard (CSH) [4, 5]. CIF curves are used to describe or explore patients' trend of survival in cases of competing risks, and the measures of treatment effect corresponding to the SDH and CSH are the sub-distribution hazard ratio (SHR) and cause-specific hazard ratio (CHR), respectively. Lau et al. [3] pointed out that the CHR regards competing events as right censored and is more suitable for epidemiologic studies, while the SHR is good at estimating risk factors and treatment effects, which makes it more applicable in clinical studies. Thus, the SHR is given as the commonly used descriptive index for the comparison of CIFs between groups. However, the SHR also has limitations in some applications: i) the most commonly used method, the Gray test [6], needs to satisfy the proportional SDH assumption [7]; ii) normally, the descriptive statistic used for competing risks data is the CIF curve; however, the statistical inference is the Gray test based on the SDH; thus, the statistical description and statistical inference do not match exactly; iii) when the SHR is used to summarize the treatment effect, the test framework contains only an SHR of the treatment group vs. the control group instead of the SDH for each group. Without baseline (control group) information, the SHR may be a relatively abstract concept for patients [8, 9]; iv) the estimation of the SDH is based on conditional probability, so that the SHR does not reflect the risk ratio of two groups, which complicates the interpretation of survival outcomes [10].

Based on the above limitations of the SHR, the median time can reflect the effect of survival, but only on a single time point, which is not a meaningful way to summarize the effect on patients over a time period. Calkins et al. [11] referred to the concept of the restricted mean survival time under competing risks (RMSTc), which is based on the restricted mean survival time (RMST) [12, 13] spent in a state free of composite events. However, the simple use of composite endpoints may not have clinical meaning [14]. In addition, the RMSTc causes the loss of accuracy regarding the event of interest, and the result can be simplified to the RMST based on the single endpoint by taking all events as one composite event.

Anderson [15] defined the number of life years lost under competing risks settings and proposed a regression model based on pseudovalue observations. Zhao et al. [16] introduced the restricted mean time lost (RMTL), which corresponds to the area under the CIF curve of the event of interest and represents the average of the lost time for the event of interest within a specific restricted period of time.

This paper develops statistical methods based on the RMTL that can avoid the limitations of the SHR and RMSTc. The paper is organized as follows. Section 2 presents the definition and estimation of measures based on the hazard, the RMTL, and the corresponding hypothesis tests and sample size formulas. In Section 3, we conduct simulation studies to assess the impact of the proposed tests. Two examples are used to illustrate the proposed methods in Section 4. Section 5 provides a discussion of our research.

## 2. Methods

Assume a randomized study with $n$ patients in two groups ($k$=1,2). The time-to-event and censoring times are denoted by $T = t_i, i = 1, 2, ..., n$ and $C$, respectively. For simplicity, we assume that $C$ is independent of $T$. $\tau$ is the truncation time point, also called the cut-off time point. Without loss of generality, two endpoints are assumed in this research: one event of interest ($j$=1) and one competing event ($j$=2). Let $I_1(t)$ and $I_2(t)$ be the CIFs for the event of interest and competing event, respectively. Based on the nonparametric maximum likelihood estimation of the CIF, the estimate of CIF $I_j(t)$ is $\hat{I}_j(t) = \sum_{t_i \leq t}(d_{ij}/n_i)\hat{S}(t_{i-1})$, where $d_{ij}$ is the number of events of type $j$ that occur at time $t_i$, the number of individuals at risk at $t_i$ is denoted by $n_i$, and $\hat{S}(t)$ is the Kaplan-Meier estimate when all events (both $j=1$ and $j=2$) are considered.

2.1 *Descriptive Analysis*

2.1.1 *Existing Measure Based on the CSH and SDH*

Both the CSH and the SDH are hazard-related measures. Under competing risks, the CSH of an event of interest is defined as

$$\lambda_{CSH}(t) = \lim_{\Delta t \to 0} \frac{P(t \leq T < t+\Delta t, j=1 \mid T \geq t)}{\Delta t}, \tag{1}$$

which indicates the patients' hazard of the event of interest at $t$ without any prior event. The corresponding descriptive measure CHR is the ratio of the CSHs of two groups.

The SDH of an event of interest is given by

$$\lambda_{SDH}(t) = \lim_{\Delta t \to 0} \frac{P(t \leq T < t+\Delta t, j=1 \mid T > t \cup (T < t \cap j \neq 1))}{\Delta t}, \tag{2}$$

which describes the patients' hazard of the event of interest at $t$ only without previously having experienced the event of interest. The corresponding descriptive measure SHR is the ratio of the SDHs of two groups.

In formulas (1) and (2), the main difference between the CSH and SDH is the number of individuals at risk. For the CSH, the number at risk at $t$ includes only patients who do not experience any type of event, while the number at risk of the SDH includes patients who do not experience the event of interest while still including patients who have experienced competing events.

### 2.1.2 *Alternative Measure Based on the RMTL*

The RMTL of the event of interest is defined as $\text{RMTL} = \int_0^\tau I_1(t)dt$ [15, 16]. Then, based on the CIF estimation of the event of interest, $\hat{I}_1(t)$, the estimate of the RMTL is given by

$$\widehat{\text{RMTL}} = \int_0^\tau \hat{I}_1(t)dt, \tag{3}$$

and the estimated variance in $\widehat{\text{RMTL}}$ is

$$\widehat{\text{var}}_{\text{RMTL}} = E(\widehat{\text{RMTL}}^2) - E(\widehat{\text{RMTL}})^2 = 2\tau \int_0^\tau \hat{I}_1(t)dt - 2\int_0^\tau t\hat{I}_1(t)dt - [\int_0^\tau \hat{I}_1(t)dt]^2, \tag{4}$$

where $E(\widehat{\text{RMTL}}^2)$ is estimated by

$$E(\text{RMTL}^2) = E[(\tau - T)^2 \mid T \leq \tau]\Pr(T \leq \tau) + E[(\tau - T)^2 \mid T > \tau]\Pr(T > \tau)$$
$$= \int_0^\tau (\tau - u)^2 f_1(u)du + 0 = 2\tau \int_0^\tau I_1(u)du - 2\int_0^\tau uI_1(u)du,$$

and $f_1(t)$ is the density function of $I_1(t)$.

The descriptive measure of the RMTL is the RMTL difference between two groups. From formula (3), the effect size of the difference in the RMTLs (RMTLd) is related to the difference between the two areas under the CIF curves.

### 2.2 *Hypothesis Test Procedures*

### 2.2.1 *Existing Test Procedures Based on the CSH and SDH*

The log-rank test can be directly used as the test corresponding to the CSH [17].

The most commonly used test for the SDH is the Gray test (Gray) [6], the test statistic of which is defined as

$$z_k = \int_0^\tau \varpi_k(t)\{\lambda_{SDH}^{(1)}(t) - \lambda_{SDH}^{(2)}(t)\}dt,$$

where $\lambda_{SDH}^{(k)}(t)$ is the estimate of the SDH for group $k$. The weight function is defined as

$\varpi_k(t) = n_k(t)\dfrac{1 - \hat{I}_k(t-)}{\hat{S}_k(t-)}$, $n_k(t)$ is the number of individuals at risk at time $t$ in group $k$,

$\hat{I}_k(t-)$ is the left-hand limit of the CIF for the event of interest in group *k*, and $\hat{S}_k(t-)$ is the left-hand limit of the probability of being free of any event in group *k*, as estimated by the Kaplan-Meier method.

2.2.2.2 *New Tests Based on the RMTLd*

*Basic Difference Test*

Assuming that $\Delta$ is the RMTLd between two groups, then the estimates of $\Delta$, $\hat{\Delta}$, are $\hat{\Delta} = \int_0^\tau [\hat{I}_{12}(t) - \hat{I}_{11}(t)]dt$, where $\hat{I}_{1k}(t)$ is the CIF estimate for the event of interest in group *k*. Thus, we present a basic test procedure based on the RMTLd. Under the null hypothesis $H_0 : \Delta = 0$, the basic difference test (Diff) statistic is given by $Z = \dfrac{\hat{\Delta}}{\sqrt{\operatorname{var}(\hat{\Delta})}} \sim N(0,1)$, and the estimate variance $\operatorname{var}(\hat{\Delta})$ is derived by the delta method; that is,

$$\operatorname{var}(\hat{\Delta}) = \operatorname{var}_1/n_1 + \operatorname{var}_2/n_2, \tag{5}$$

where $\operatorname{var}_k = 2\tau \int_0^\tau \hat{I}_{1k}(t)dt - 2\int_0^\tau t\hat{I}_{1k}(t)dt - [\int_0^\tau \hat{I}_{1k}(t)dt]^2$ according to formula (3), and $n_k$ is the sample size in the *k*th group.

*Supremum Difference Test*

We refer to the supremum difference test (sDiff) statistics [18] based on the RMTLd. The test statistic is given by $Q_S = \sup\{|\hat{\Delta}(t_r)|, t_r \leq \tau\} / \hat{\sigma}(\tau)$, where $\hat{\Delta}(t_r)$ is calculated by

$$\hat{\Delta}(t_r) = \sum_{t_i \leq t_r}[\hat{I}_{12}(t_i) - \hat{I}_{11}(t_i)](t_{i+1} - t_i),\ t_r = t_1, t_2, ..., \tau. \tag{6}$$

The standard error of $\hat{\Delta}(t_r)$ is solved by

$$\sigma^2(\tau) = \sum_{i|t_i<\tau} (t_{i+1}-t_i)^2 \{\hat{V}[\hat{I}_{12}(t_i)] + \hat{V}[\hat{I}_{11}(t_i)]\}$$

$$+ \sum_{i<i'|t_i,t_{i'}<\tau} 2\rho(t_{i+1}-t_i)(t_{i'+1}-t_{i'}) \times \{\hat{V}[\hat{I}_{12}(t_i)]+\hat{V}[\hat{I}_{11}(t_i)]\}\{\hat{V}[\hat{I}_{12}(t_{i'})]+\hat{V}[\hat{I}_{11}(t_{i'})]\}\}^{1/2}$$

based on Aalen's variance [19] of the CIF estimator, where $\rho$ is the correlation coefficient between $\hat{I}_{12}(t_i)-\hat{I}_{11}(t_i)$, and $\hat{I}_{12}(t_{i'})-\hat{I}_{11}(t_{i'})$, where $i \neq i'$. $\rho$ is difficult to estimate because it involves the assumption of an unknown underlying CIF distribution of the actual data. Lyu et al. [20] found that when $\rho=0.50$, the test statistic does not inflate the type I error rate and maintains high power. Hence, we fixed $\rho$ at an acceptable value of 0.5 in this article.

Under the null hypothesis, the distribution of $Q_S$ can be approximated by the distribution of $\sup\{|M(x)|, 0 \leq x \leq 1\}$, where $M$ is a standard Brownian motion process. According to Billingsley [21], the probability distribution of $\sup|M(t)|$ is given by

$$P[\sup|M(t)|>x] = 1 - \frac{4}{\pi}\sum_{a=0}^{\infty}\frac{(-1)^a}{2a+1}e^{[-\pi^2(2a+1)^2/8x^2]} . \qquad (7)$$

Assuming that formula (7) converges when $a \to m$ [22], then $m$ is solved as $m = \max\left\{\left\lceil \frac{x\sqrt{2}}{\pi}\sqrt{\log\frac{1}{\pi\varepsilon}} - \frac{1}{2}\right\rceil, 1\right\}$, where $\lceil \cdot \rceil$ refers to the minimum positive integer of this value, and $\varepsilon$ is the permissible error for estimating $P[\sup|M(t)|>x]$.

2.3 *The Sample Size Formula Based on the RMTLd*

Under competing risks, the use of Gray is limited by the proportional SDH assumption. Hence, the corresponding sample size formula is not always available. In addition, the estimated Gray sample size (based on the Gray) depends on the incidence of the event of interest; that is, a large deviation between the actual incidence and the assumed incidence results in a broad range of estimated sample sizes. This paper does not discuss the sample size formula for the RMSTc (which can be estimated by the method based on the single endpoint), as our focus is on the event of interest. The sample size formulas based on Diff and sDiff are proposed in the following section.

2.3.1 *Method Based on the Basic Difference Test*

According to Diff, as shown in Section 2.2.2, the following hypotheses are considered: $H_0: \Delta = 0;\ H_1: \Delta \neq 0$. The test statistic under the null hypothesis $Z = \dfrac{\hat{\Delta}}{\sqrt{\sigma(\hat{\Delta})}} \sim N(0,1)$ is rejected at the approximate $\alpha$ level of significance if $\left|\dfrac{\hat{\Delta}}{\sqrt{\sigma(\hat{\Delta})}}\right| > z_{\alpha/2}$. We write the expression $z_{1-\alpha/2} = \Phi^{-1}(1-\alpha/2)$, where $\Phi$ is the standard normal distribution. Then, under the alternative hypothesis with a desired power of $1-\beta$, the sample size can be obtained by solving $1-\beta = P(|Z| > z_{1-\alpha/2} | H_1) = P(Z > z_{1-\alpha/2} | H_1) + P(Z < -z_{1-\alpha/2} | H_1)$.

By symmetry of the normal distribution, $P(|Z| > z_{1-\alpha/2} | H_1)$ and $P(Z < -z_{1-\alpha/2} | H_1)$ are equal. Thus, we obtain $1-\beta \simeq \Phi(-z_{1-\alpha/2} + \dfrac{\Delta}{\sigma})$, i.e., $(z_{1-\beta} + z_{1-\alpha/2})^2 \sigma = \Delta$.

According to formula (5), we have $\sigma^2 = \dfrac{\sigma_1^2}{n_1} + \dfrac{\sigma_2^2}{n_2} = (1+r) \times (\dfrac{\sigma_1^2}{n} + r^{-1} \times \dfrac{\sigma_2^2}{n})$, where $\dfrac{n_2}{n_1} = r$.

Thus, the total required sample size is

$$n = (1+r)\dfrac{(z_{1-\beta} + z_{1-\alpha/2})^2}{\Delta^2 / (\sigma_1^2 + r^{-1}\sigma_2^2)}.$$

2.3.2 *Method Based on the Supremum Difference Test*

In the sample size calculation of the supremum test, the main purpose is to obtain $\xi$ in function $n = \xi \cdot \tilde{n}$, where $\tilde{n}$ is the calculated sample size based on Diff.

As with $H_0: \Delta = 0;\ H_1: \Delta \neq 0$, we assume that $\Delta = \eta \neq 0$ under the alternative hypothesis. Then, we write the expression

$$U(t) = \frac{\hat{\Delta}(t)}{\hat{\sigma}(\tau)} = M(u(t)) + \eta u(t),$$

where $M(\cdot)$ is a standard Brownian motion process and $u(t)$ is a time function. Then, $U(t)$ is a standard Brownian motion process that deviates with a mean of $\eta$. Here, we assume $\eta = \lambda\sqrt{R}$ with a fixed effect size $\lambda$, where $R = nR(\tau)$, and $R(t)$ is the probability that the event of interest happened before $t$. Then, the relation of $R$ and $\eta$ is given by

$$R = \frac{\eta^2}{\lambda^2}. \tag{8}$$

Assume that $V_{1-\alpha/2}$ is the critical value of the supremum value of the standard Brownian motion process, i.e.,

$$P\{\sup_{u\in[0,1]} |M(u)| > V_{1-\alpha/2}\} = P\{\sup_{u\in[0,1]} M(u) > V_{1-\alpha/2}\} + P\{\inf_{u\in[0,1]} M(u) < -V_{1-\alpha/2}\} = \alpha. \tag{9}$$

By the symmetry of Brownian motion, both probabilities, $P\{\sup_{u\in[0,1]} M(u) > V_{1-\alpha/2}\}$ and $P\{\inf_{u\in[0,1]} M(u) < -V_{1-\alpha/2}\}$, are equal. Hence, we only need to consider the calculation of $P\{\sup_{u\in[0,1]} M(u) > V_{1-\alpha/2}\}$ here. According to the joint distribution of $\sup_{u\in[0,1]} M_\eta(u)$ and $M_\eta(1)$ [22], $M_\eta(\cdot)$ is the Brownian motion with a mean of $\eta$. Eng and Kosorok [23] obtained the following function after integration:

$$P\{\sup_{u\in[0,1]} M_\eta(u) > x\} = \bar{\Phi}(x-\eta) + e^{2\eta V_{1-\alpha/2}} \bar{\Phi}(x+\eta).$$

Thus, the sample size needed to achieve a desired power of $1-\beta$ with a two-sided type I error of $\alpha$ can be obtained by

$$\bar{\Phi}(V_{1-\alpha/2} - \eta) + e^{2\eta T_{1-\alpha/2}} \bar{\Phi}(V_{1-\alpha/2} + \eta) = 1 - \beta, \tag{10}$$

where $\bar{\Phi} = 1 - \Phi$ and $\Phi$ is the standard normal distribution.

Under the alternative hypothesis, we obtain the limiting distribution of $U_n(\tau)$ as $M_\eta(1)$, which is a normal deviation with mean $\eta$ and variance 1. With a critical value $Z_{1-\alpha/2}$, we solve for $\tilde{\eta}$ in the following expression: $\bar{\Phi}(Z_{1-\alpha/2} - \tilde{\eta}) = 1 - \beta$, that is, $\tilde{\eta} = Z_{1-\alpha/2} + Z_{1-\beta}$. Hence, we obtain

$$\tilde{R} = \frac{(Z_{1-\alpha/2} + Z_{1-\beta})^2}{\lambda^2}. \tag{11}$$

Because formulas (8) and (11) have the same effect size $\lambda$, the denominators cancel, and the ratio becomes $\frac{R}{\tilde{R}} = \frac{\eta^2}{\tilde{\eta}^2} = \xi$, where only $\eta$ remains to be solved.

First, we need to estimate the critical value $V$ in formula (9). From the cumulative probability distribution [21]

$$P[\sup |M(t)| \leq x] = \frac{4}{\pi} \sum_{a=0}^{\infty} \frac{(-1)^a}{2a+1} e^{[-\pi^2(2a+1)^2/8x^2]},$$

let $V = 1 - \frac{4}{\pi} \sum_{a=0}^{\infty} \frac{(-1)^a}{2a+1} e^{[-\pi^2(2a+1)^2/8x^2]}$, and assume that $V$ converges when $a \to m$. Then, $m$ is solved as $m = \max\left\{\left\lceil \frac{x\sqrt{2}}{\pi} \sqrt{\log \frac{1}{\pi\varepsilon}} - \frac{1}{2} \right\rceil, 1\right\}$ [23], where $\lceil \cdot \rceil$ refers to the minimum positive integer of this value, and $\varepsilon$ represents the residuals. Finally, assume a function, $Y(x) = \bar{\Phi}(u-x) + e^{2ux}\bar{\Phi}(u+x)$, with a derivative function of $Y'(x) = -\bar{\varphi}(u-x) + e^{2ux}\bar{\varphi}(u+x) + 2ue^{2ux}\bar{\Phi}(u+x)$. Use the Newton-Raphson iteration to solve for $\eta$ in formula (10), let $o_i = \frac{(1-\beta) - Y(x)}{Y'(x)}$, and iterate the cycle until $o \leq \varepsilon$; the estimate $\eta$ is given by $\eta = \tilde{\eta} + \sum o_i$.

## 3. Results

3.1 *Hypothesis Test Procedures*

### 3.1.1 *Simulation Design*

To compare the performance of the above tests, Monte Carlo simulations were carried out to study the type I error and the statistical power under a variety of situations. The following procedures were performed to test the hypotheses: Gray, Diff, and sDiff. The performance of these tests was evaluated by using 6 alternative scenarios (Figure 1): (A) two groups with no difference (the comparison for type I error); (B) two groups with a proportional SDH difference; (C) two groups with a non-proportional SDH difference; (D) an early difference in the CIFs; (E) CIFs with a late difference; (F) two CIFs with a cross difference.

Let $\tau_1$ and $\tau_2$ be the last event of interest time in the two groups. Here, we considered a commonly used option, the minimum of the last event in the time of interest in two groups ($\tau = \min(\tau_1, \tau_2)$), as $\tau$. The event of interest and the competing event were generated from CIFs with piecewise Weibull distributions. The specific parameter settings are presented in Web Table A1. The distribution of events was based on the binomial distribution $B(N, p_1)$, where $N$ represents the sample size of each group and $p_1$ is the maximum cumulative incidence of events of interest, which is set as $p_1 = I_1(\infty) = 0.7$. The censored times $C$ in the two groups were generated from uniform distributions. Then, each individual was assigned an observed time $t=\min(T, C)$ and the event indicator $\delta=0[T>C]$. By changing the distribution parameters of $C$, both groups were set to have the same censoring rates of approximately 0, 15%, 30% and 45%. We also considered equal group sizes ($n_1=n_2=50, 100, 150$) and unequal group sizes ($n_1=50, n_2=100; n_1=50, n_2=150; n_1=50, n_2=200$). All simulations were performed using 5000 iterations. The nominal significance level of each method was fixed at the conventional level of 0.05.

3.1.2 *Simulation Results*

As Table 1 shows, the type I error rates for Diff are stable under 0 censoring but gradually inflate with increasing censoring rates, which represent the most radical test. As the type I error rates of Diff are inflated, this test is not included in the comparison of test power. Compared to Diff, Gray is steadier. The type I error rates of sDiff are relatively conservative for light censoring but increase with increasing censoring rates.

The power results are shown in Table 1. When two CIF curves have a proportional SDH (Figure 1B), the powers of all the tests increase with increasing sample size but decline with increasing censoring rates. Gray demonstrates the optimal power in this situation, followed by sDiff. For the non-proportional SDH difference (Figure 1C), sDiff is the most powerful test, while Gray has the lowest power in this situation. For the early difference in the CIF curves (Figure 1D), with increasing censoring rates, sDiff becomes much more powerful, and Gray exhibits the lowest power in this situation. When considering the late difference in the CIF curves (Figure 1E), the powers of all tests decline with increasing censoring rates. In this situation, Gray is more powerful, followed by sDiff. In the case of a cross difference in the CIF curves (Figure 1F), Gray has the lowest power. With increasing censoring rates, sDiff is much more powerful than Gray.

Note that in situation C (non-proportional SDH), situation D (early difference), and situation F (cross difference), all tests exhibit gradually increasing power with increasing censoring rates. This result occurs because the two CIF curves are not convergent in the later period but diverge with the increased censoring, which makes the increased difference between the two CIF curves proportional.

To summarize the simulation results, we applied the analysis of variance (ANOVA) technique [24] to evaluate the type I error and power. For type I error, the absolute small and close-to-zero estimates indicate that rates are close to 0.05. For power, good performance is indicated by large estimated values (see details in Appendix A). Table 2 shows that sDiff corrects the inflated type I error of Diff when censoring occurs. In Table 3, sDiff is slightly lower than Gray when there is a proportional SDH (situation B), and when there is a late difference (situation E), the difference between Gray and sDiff is approximately only 2.242%, whereas the powers of sDiff are much higher than those of Gray in other situations. Considering all situations, combinations of sample sizes, and censoring rates, sDiff performs better in most situations.

3.2 *Calculations of Sample Size*

3.2.1 *Simulation Design*

A simulation study was also performed to evaluate the proposed sample size formula. Two scenarios were considered (Figure 1 B-C): (B) two groups with a proportional SDH difference and (C) two groups with a non-proportional SDH difference. Both scenarios were examined under four scenarios with either 0.05 or 0.01 for the two-sided type I error and with either 0.8 or 0.9 as the power. The follow-up time, $\tau$, which is also the truncation time point, was set as the minimum of the last observed time of the pilot study for two groups. Assume two groups with an equal sample size, i.e., $r=1$. Then, based on situation B and situation C, the necessary parameters were estimated by simulation, and finally, we obtained the calculated sample size with the given parameters. In addition, Monte Carlo simulations were used to examine the observed power. The simulations were performed using 1000 replications.

3.2.2 *Simulation Results*

As shown in Table 4, the calculated sample size of all the tests increases with a decreasing type I error rate and with an increasing target power. When the CIFs satisfy the proportional SDH assumption (situation B, Figure 1B), the calculated sample sizes of Gray, Diff and sDiff are close to each other, with Diff having the highest observed power. In this situation, Gray and sDiff have a similar observed power, which is close to the target power. When there is a non-proportional SDH (situation C, Figure 1C), the calculated sample sizes of Gray are much higher than those of Diff and sDiff, and sDiff has a relatively high observed power. In addition, the observed powers of Gray do not reach the target power in this situation.

In addition, the comparison of power for Gray, Diff and sDiff with a fixed sample size (calculated by sDiff) is shown in Web Table A2. The results show that the power under situation B for Diff is larger than that for Gray and sDiff, but the three tests have similar power. However, in situation C, the power of Gray is much lower than that of Diff and sDiff.

As the type I error rates of Diff are inflated with censoring, the corresponding sample size formula is considered unstable. In general, when the CIFs satisfy the proportional SDH assumption, both Gray and sDiff can be considered; when the SDH is non-proportional, sDiff is considered more adaptive.

## 4. Applications

4.1 *Example 1: Bone Marrow Transplantation Data*

The data used to evaluate the effect of T-cell depletion on bone marrow transplantation [25] came from 408 patients divided into a T-cell depleted group (Yes) with 354 cases and a T-cell not depleted group (No) with 54 cases. The censoring rates for the two groups were approximately 41% and 28%, respectively. The study included two types of events: death from treatment-related causes, which was defined as the event of interest, and relapse, which was set as a competing event. At the end of follow-up, 161 patients (146 from the Yes group and 15 from the No group) experienced an event of interest, and 87 patients (70 from the Yes group and 17 from the No group) experienced competing events. A test of the proportional SDH assumption yielded a result of *P*=0.264.

The descriptive statistics and the hypothesis test results for the examples are shown in Table 5. In the hazard-related measures, the CHR and SHR showed that the ratios of the Yes group vs. the No group were 0.86 (0.59, 1.25) and 0.60 (0.36, 1.00), respectively. However, the log-rank test, which is based on the CHR, showed no significant differences (*P*=0.053), while Gray based on the SHR indicated that there were significant differences between the two groups (*P*=0.049). In addition, we could not obtain the estimated CSH or SDH for either group, which led to a lack of descriptive information for either group; only a CHR or an SHR could be obtained. This outcome led to difficulty in clinical interpretation.

For the composite endpoint, the RMSTc showed that the mean survival time of the patients in the Yes group was 1.83 (-5.03, 8.69) months longer than that of the patients in the No group within the truncation time point of 41.8 months, and there were no significant differences (*P*=0.601). Additionally, the RMSTc could not provide information regarding treatment-related death.

Let $\tau$ =41.8 months; Table 5 shows that the RMTL of treatment-related death in the Yes group was 9.57 (5.18, 13.96) months, which corresponds to the area under the CIF curve, i.e., S2 in Figure 2A. In the No group, the RMTL corresponds to the area under the CIF curve, i.e., S1+S2=15.49 (13.53, 17.45) months. Hence, the difference in RMTL between the two groups has an area of S1, which means that the patients in the Yes group took 5.92 (1.11, 10.72) months longer to succumb to treatment-related death. According to Table 5, the RMTL-based tests (Diff and sDiff) showed significance at the conventional level of 0.05.

As shown in Figure 2B, a selection of different $\tau$ values led to a difference in the calculated sample size for Diff and sDiff: the calculated sample size increased with increasing $\tau$ and became steady after 20 months. The calculated sample sizes at $\tau = 41.8$ months were 280 and 298 for Diff and sDiff, respectively, both of which were close to the sample size calculated by Gray (*n*=300).

### 4.2 *Example 2: Lymphocytic Leukemia Data*

A previous study compared the effects of radiotherapy in the treatment of patients with lymphocytic leukemia (LL). A total of 1400 patients were randomly extracted from the Surveillance, Epidemiology, End Results (SEER) Program. Among these patients, two groups were included: the no radiotherapy group (No RT) consisted of 1318 cases, and the radiotherapy group consisted of 82 cases. The censoring rates in the two groups were approximately 3% and 44%, respectively. During the follow-up, death from LL was set as the event of interest; death from other causes was recorded as a competing risk. At the end of the research, 364 patients (333 from the No RT group and 31 from the RT group) had died from LL, and 467 patients (452 from the No RT group and 15 from the RT group) had died from other causes. The corresponding test indicated a severe violation of the proportional SDH assumption ($P$=0.006).

Regarding the hazard-related indexes, Table 5 shows that the No RT group had a lower hazard ratio than the RT group (CHR=1.14 (0.78,1.65); SHR=1.45 (0.98, 2.14)). However, in this example, the SHR varied with time ($P$=0.006) instead of being constant. Therefore, the CHR and SHR may not be available for this example.

When considering the composite endpoint, the RMSTc showed that the mean survival time of the RT patients was 0.66 (-1.13,2.28) years longer than that of the No RT patients within the truncation time point of 15.3 years (Table 5), which reflected the overall survival but could not reflect the survival rates of patients who died of LL.

Let $\tau = 15.3$ years; Table 5 shows that the RMTL of LL-related death in the RT group was 4.68 (3.34, 6.03) years, which is equal to the area of S1+S2 in Figure 2C and corresponds to the area under the CIF curve. In the No RT group, the RMTL was S2=2.96 (2.69, 3.24) years. The difference in the RMTL between the two groups was S1=1.72 (0.35, 3.09) years, which is the delay time until the patients in the No RT group succumbed from LL-related death. As shown in Table 5, for all test procedures, only Diff and sDiff, which were based on the RMTL, showed significance at the conventional level of 0.05.

As Figure 2D shows, with increasing $\tau$, the calculated sample sizes showed a trend of decreasing first and then increasing, and they reached the smallest estimation of sample size at approximately 7 years, which was much less than the results found with Gray (*n*=886). The calculated sample sizes at $\tau = 15.3$ years were 344 and 364 for Diff and sDiff, respectively.

## 5. Discussion

When dealing with competing risks datasets, the SHR is often used as a typical descriptive method with the test procedures. However, because the SHR lacks baseline information (a control group) and does not directly reflect the risk ratio of the two groups, it may complicate the interpretation of the survival outcome and may be a relatively abstract concept for patients. The RMSTc can directly describe patient survival and does not depend on the proportional SDH assumption, but the simple use of composite endpoints does not always have clinical meaning and degrades the accuracy of patient information [14]. Conversely, the RMTL can avoid some limitations of the above methods. Moreover, the relationship between the RMTL and RMSTc can be derived as $\text{RMTL}_1 + \text{RMTL}_2 + \cdots + \text{RMTL}_j + \text{RMST}_c = \tau$, where RMTL$_j$ means the area under the CIF curves for cause *j*. As RMTL$_j$ is interpreted as the average survival time lost due to cause *j* within $\tau$, the RMTL can be observed from the CIF curves directly, while the SHR cannot directly reflect the CIF curves. In addition, Gray, which corresponds to the SHR, needs to satisfy the proportional SDH assumption, while the RMTL-related tests do not. From the simulation results of the hypothesis testing procedures, sDiff, which is based on the RMTLd, corrects the severe skewness of Diff under high censoring and has improved power under various scenarios compared to Gray. In general, sDiff maintains good performance compared to Gray and Diff. In addition, this paper also contains sample size formulas based on the RMTLd. When the proportional SDH assumption is satisfied, the calculated sample sizes of Diff and sDiff are close to that of Gray, while Diff still has the highest power. Because the type I error rates of Diff are inflated with censoring, we still suggest that Gray and sDiff be used in this situation; however, when the SDH is non-proportional, the sample sizes estimated by Gray are much larger than those estimated by Diff and sDiff with the lowest observed power. Hence, in this situation, sDiff seems more adaptable for use.

The sample sizes calculated in the examples (Figure 2B, D) suggest that different choices of $\tau$ may have a large influence on the calculation of the sample size. In example 1, the calculated sample size increases with increasing $\tau$ and is similar to the sample size estimated by Gray after 20 months (Figure 2B), while in example 2, the calculated sample sizes show a trend of decreasing first and then increasing (Figure 2D). Hence, it is important to choose an appropriate $\tau$ for the calculated sample size of Diff and sDiff. In practical research, $\tau$ is always determined as the follow-up time in the study design. If all patients in one of the groups experience an endpoint during the follow-up period, then the study is stopped, and this time point is determined as the final analysis of the study, i.e., $\tau$; otherwise, if patients in either group do not have a completely observed endpoint until the end of the follow-up period, then the designed follow-up period is regarded as the truncation time point. In this paper, the calculated sample sizes in simulations are based on the minimum time of the last observation of the event of interest in the two groups as $\tau$. The issue of how to define an appropriate $\tau$ in a specific research context will be considered in a future study.

## 6. Conclusions

The RMTL can meaningfully summarize treatment effects for clinical decision making, which can be reported with the SDH ratio for competing risks data. The proposed sDiff test is robust and can be considered for statistical inference in real data analysis; the two proposed calculated sample size formulas have wide applicability and can also be applied to real trial designs.

**Additional file**

The additional file contains the theory of variance techniques (ANOVA) used to evaluate the type I error and power, the parameter settings of the two CIFs for the simulations and the comparison of power with a fixed sample size.

**Competing interests**

The authors declare that they have no competing interests.

**Funding**

This work is supported by the National Natural Science Foundation of China (81673268, 81903411), Natural Science Foundation of Guangdong Province (2018A030313849) and the Guangdong Basic and Applied Basic Research Foundation (2019A1515011506).

# References

1. Bakoyannis G, Touloumi G. Practical methods for competing risks data: a review. Stat Methods Med Res.2012; 21: 257-272.

2. Van WC, Mcalister FA. Competing risk bias was common in Kaplan-Meier risk estimates published in prominent medical journals. J Clin Epidemiol. 2016; 69: 170-173.

3. Lau B, Cole SR, Gange SJ. Competing risk regression models for epidemiologic data. Am J Epidemiol. 2009; 170: 244-256.

4. Kalbfleisch JD, Prentice RL. The statistical analysis of failure time data (second edition). John Wiley & Sons. 2002.

5. Fine JP, Gray RJ. A proportional hazards model for the subdistribution of a competing risk. J AM Stat Assoc. 1999; 94: 496-509.

6. Gray RJ. A class of K-sample tests for comparing the cumulative incidence of a competing risk. Ann Stat. 1988; 16: 1141–1154.

7. Li J, Scheike TH, Zhang MJ. Checking Fine and Gray subdistribution hazards model with cumulative sums of residuals. Lifetime Data Anal. 2015; 21: 197-217.

8. Uno H, Wittes J, Fu H, Solomon SD, Claggett B, Tian L, et al. Alternatives to hazard ratios for comparing the efficacy or safety of therapies in noninferiority studies. Ann Intern Med. 2015; 163: 127-134.

9. Kim DH, Uno H, Wei LJ. Restricted mean survival time as a measure to interpret clinical trial results. JAMA Cardiol. 2017; 2: 1179-1180.

10. Sutradhar R, Austin PC. Relative rates not relative risks: addressing a widespread misinterpretation of hazard ratios. Ann Epidemiol. 2018; 28: 54-57.

11. Calkins KL, Canan CE, Moore RD, Lesko CR, Lau B. An application of restricted mean survival time in a competing risks setting: comparing time to ART initiation by injection drug use. BMC Med Res Methodol. 2018; 18: 27.

12. Royston P, Parmar MKB. The use of restricted mean survival time to estimate the treatment effect in randomized clinical trials when the proportional hazards assumption is in doubt. Stat Med. 2011; 30: 2409–2421.

13. Royston P, Parmar MKB. Restricted mean survival time: an alternative to the hazard ratio

for the design and analysis of randomized trials with a time-to-event outcome. BMC Med Res Methodol. 2013; 13: 152.

14. Mell LK, Jeong JH. Pitfalls of using composite primary end points in the presence of competing risks. J Clin Oncol. 2010; 28: 4297-4299.

15. Andersen PK. Decomposition of number of life years lost according to causes of death. Stat Med. 2013; 32: 5278-5285.

16. Zhao L, Tian L, Claggett B, Pfeffer M, Kim DH, Solomon S, et al. Estimating treatment effect with clinical interpretation from a comparative clinical trial with an end point subject to competing risks. JAMA Cardiol. 2018; 3: 357-358.

17. Tai BC, Wee J, Machin D. Analysis and design of randomised clinical trials involving competing risks endpoints. Trials. 2011; 12: 127.

18. Gill RD. Censoring and stochastic integrals. Amsterdam: The Mathematical Center. 1980.

19. Aalen O. Nonparametric estimation of partial transition probabilities in multiple decrement models. Ann Stat. 1978; 6: 534-545.

20. Lyu J, Chen J, Hou Y, and Chen Z. Comparison of two treatments in the presence of competing risks. Pharmaceutical Statistics. 2020. DOI: 10.1002/pst.2028

21. Billingsley P. Converge of probability measures. New York: Wiley. 1968.

22. Borodin AN, Salminen P. Handbook of brownian motion-facts and formulae. Basel: Birkhauser. 2000.

23. Eng KH, Kosorok MR. A sample size formula for the supremum log-rank statistic. Biometrics. 2005; 61: 86-91.

24. Klein JP, Logan B, Harhoff M, et al. Analyzing survival curves at a fixed point in time. Stat Med. 2007; 26(24):4505.

25. Thomas HS. timereg: flexible regression models for survival data. R package version 1.9.3. 2019. https://CRAN.R-project.org/package=timereg.

**Table 1** *Type I error rates and powers of the test procedures*

| $n_1, n_2$ | Cen (%) | Situation A | | | Situation B | | | Situation C | | | Situation D | | | Situation E | | | Situation F | | |
|---|---|---|---|---|---|---|---|---|---|---|---|---|---|---|---|---|---|---|---|
| | | Gray | Diff | sDiff | Gray | Diff | sDiff | Gray | Diff | sDiff | Gray | Diff | sDiff | Gray | Diff | sDiff | Gray | Diff | sDiff |
| 50,50 | 0 | 0.046 | 0.053 | 0.025 | 0.795 | 0.813 | 0.689 | 0.128 | 0.383 | 0.109 | 0.076 | 0.258 | 0.155 | 0.251 | 0.228 | 0.182 | 0.053 | 0.190 | 0.153 |
| | 15 | 0.049 | 0.063 | 0.033 | 0.784 | 0.799 | 0.678 | 0.200 | 0.438 | 0.161 | 0.101 | 0.307 | 0.185 | 0.207 | 0.171 | 0.134 | 0.070 | 0.234 | 0.192 |
| | 30 | 0.049 | 0.074 | 0.043 | 0.707 | 0.734 | 0.639 | 0.302 | 0.555 | 0.329 | 0.142 | 0.381 | 0.247 | 0.129 | 0.100 | 0.076 | 0.105 | 0.328 | 0.288 |
| | 45 | 0.051 | 0.078 | 0.052 | 0.635 | 0.666 | 0.573 | 0.477 | 0.761 | 0.671 | 0.214 | 0.549 | 0.427 | 0.066 | 0.068 | 0.059 | 0.196 | 0.611 | 0.583 |
| 100,100 | 0 | 0.056 | 0.056 | 0.032 | 0.976 | 0.982 | 0.952 | 0.189 | 0.541 | 0.238 | 0.112 | 0.370 | 0.255 | 0.404 | 0.437 | 0.393 | 0.070 | 0.247 | 0.253 |
| | 15 | 0.052 | 0.066 | 0.032 | 0.974 | 0.980 | 0.946 | 0.310 | 0.599 | 0.288 | 0.156 | 0.423 | 0.289 | 0.364 | 0.357 | 0.307 | 0.107 | 0.289 | 0.309 |
| | 30 | 0.052 | 0.078 | 0.043 | 0.947 | 0.960 | 0.922 | 0.493 | 0.735 | 0.509 | 0.237 | 0.513 | 0.375 | 0.222 | 0.182 | 0.153 | 0.171 | 0.386 | 0.432 |
| | 45 | 0.050 | 0.082 | 0.051 | 0.907 | 0.926 | 0.885 | 0.744 | 0.931 | 0.880 | 0.375 | 0.711 | 0.632 | 0.103 | 0.100 | 0.086 | 0.332 | 0.679 | 0.782 |
| 150,150 | 0 | 0.052 | 0.055 | 0.028 | 0.997 | 0.999 | 0.994 | 0.261 | 0.669 | 0.371 | 0.136 | 0.483 | 0.359 | 0.514 | 0.611 | 0.567 | 0.078 | 0.315 | 0.429 |
| | 15 | 0.054 | 0.065 | 0.032 | 0.998 | 0.998 | 0.994 | 0.432 | 0.728 | 0.433 | 0.210 | 0.524 | 0.396 | 0.493 | 0.516 | 0.463 | 0.131 | 0.361 | 0.482 |
| | 30 | 0.052 | 0.072 | 0.040 | 0.994 | 0.995 | 0.988 | 0.660 | 0.851 | 0.682 | 0.330 | 0.610 | 0.498 | 0.298 | 0.259 | 0.223 | 0.237 | 0.472 | 0.627 |
| | 45 | 0.050 | 0.077 | 0.042 | 0.984 | 0.986 | 0.978 | 0.902 | 0.983 | 0.961 | 0.510 | 0.807 | 0.801 | 0.122 | 0.111 | 0.096 | 0.456 | 0.764 | 0.935 |
| 50,100 | 0 | 0.055 | 0.060 | 0.034 | 0.882 | 0.910 | 0.824 | 0.119 | 0.457 | 0.177 | 0.111 | 0.285 | 0.181 | 0.283 | 0.314 | 0.288 | 0.038 | 0.243 | 0.245 |
| | 15 | 0.058 | 0.071 | 0.041 | 0.877 | 0.905 | 0.824 | 0.190 | 0.519 | 0.233 | 0.152 | 0.328 | 0.216 | 0.271 | 0.262 | 0.236 | 0.058 | 0.304 | 0.294 |
| | 30 | 0.054 | 0.078 | 0.047 | 0.829 | 0.863 | 0.793 | 0.303 | 0.640 | 0.425 | 0.216 | 0.405 | 0.277 | 0.172 | 0.142 | 0.125 | 0.096 | 0.411 | 0.394 |
| | 45 | 0.050 | 0.079 | 0.050 | 0.768 | 0.808 | 0.748 | 0.521 | 0.865 | 0.783 | 0.320 | 0.588 | 0.488 | 0.085 | 0.081 | 0.074 | 0.188 | 0.663 | 0.668 |
| 50,150 | 0 | 0.050 | 0.056 | 0.029 | 0.908 | 0.931 | 0.860 | 0.114 | 0.509 | 0.203 | 0.125 | 0.292 | 0.189 | 0.297 | 0.350 | 0.333 | 0.033 | 0.288 | 0.315 |
| | 15 | 0.054 | 0.069 | 0.032 | 0.905 | 0.931 | 0.864 | 0.192 | 0.566 | 0.261 | 0.175 | 0.332 | 0.219 | 0.300 | 0.298 | 0.281 | 0.049 | 0.349 | 0.365 |
| | 30 | 0.055 | 0.078 | 0.045 | 0.869 | 0.900 | 0.846 | 0.313 | 0.701 | 0.498 | 0.252 | 0.415 | 0.291 | 0.202 | 0.164 | 0.156 | 0.085 | 0.470 | 0.475 |
| | 45 | 0.049 | 0.078 | 0.049 | 0.812 | 0.856 | 0.804 | 0.558 | 0.899 | 0.842 | 0.378 | 0.605 | 0.522 | 0.088 | 0.084 | 0.082 | 0.186 | 0.713 | 0.743 |
| 50,200 | 0 | 0.053 | 0.059 | 0.035 | 0.917 | 0.947 | 0.881 | 0.111 | 0.536 | 0.212 | 0.133 | 0.302 | 0.200 | 0.301 | 0.377 | 0.365 | 0.029 | 0.309 | 0.358 |
| | 15 | 0.056 | 0.071 | 0.039 | 0.917 | 0.944 | 0.883 | 0.182 | 0.589 | 0.279 | 0.189 | 0.346 | 0.231 | 0.313 | 0.329 | 0.324 | 0.047 | 0.377 | 0.411 |
| | 30 | 0.054 | 0.080 | 0.050 | 0.887 | 0.917 | 0.872 | 0.312 | 0.715 | 0.520 | 0.276 | 0.426 | 0.315 | 0.222 | 0.186 | 0.181 | 0.084 | 0.501 | 0.519 |
| | 45 | 0.051 | 0.082 | 0.052 | 0.841 | 0.877 | 0.833 | 0.566 | 0.917 | 0.866 | 0.409 | 0.608 | 0.543 | 0.095 | 0.098 | 0.094 | 0.181 | 0.747 | 0.780 |

Cen: censoring rate for each group.

**Table 2** *Average deviations (%) from the nominal 5% level of the tests (TEST) adjusted using ANOVA*

|  |  | Gray | Diff | sDiff |
|---|---|---|---|---|
| Model 1 | 50,50 | -0.100 | 1.735 | -1.160 |
| $n_1, n_2$ | 100,100 | 0.230 | 2.035 | -1.065 |
|  | 150,150 | 0.185 | 1.715 | -1.435 |
|  | 50,100 | 0.415 | 2.205 | -0.710 |
|  | 50,150 | 0.175 | 2.010 | -1.135 |
|  | 50,200 | 0.360 | 2.295 | -0.595 |
| Model 2 | 0 | 0.193 | 0.657 | -1.933 |
| cen | 15% | 0.373 | 1.733 | -1.533 |
|  | 30% | 0.260 | 2.673 | -0.527 |
|  | 45% | 0.017 | 2.933 | -0.073 |
| Model 4 |  | 0.211 | 1.999 | -1.017 |

cen: censoring rate.

**Table 3** *Average rejection rates for the tests (TEST) adjusted using ANOVA*

|  |  | Gray | Diff | sDiff |
|---|---|---|---|---|
| Model 1 | 50,50 | 23.184 | 37.869 | 27.650 |
| $n_1, n_2$ | 100,100 | 35.978 | 51.732 | 44.426 |
|  | 150,150 | 43.716 | 60.222 | 56.389 |
|  | 50,100 | 27.389 | 44.954 | 36.473 |
|  | 50,150 | 29.203 | 48.274 | 40.747 |
|  | 50,200 | 30.059 | 50.240 | 43.338 |
| Model 2 | 0 | 31.467 | 48.584 | 39.116 |
| cen | 15% | 34.505 | 50.335 | 40.591 |
|  | 30% | 36.983 | 53.062 | 45.576 |
|  | 45% | 43.398 | 63.546 | 60.732 |
| Model 3 | B | 87.956 | 90.108 | 84.463 |
| sit | C | 35.752 | 67.038 | 45.554 |
|  | D | 22.228 | 45.288 | 34.544 |
|  | E | 24.175 | 24.266 | 21.993 |
|  | F | 12.831 | 42.710 | 45.965 |
| Model 4 |  | 36.588 | 53.882 | 46.504 |

cen: censoring rate;

sit: simulated situation.

**Table 4** *Simulation results for sample size*

| $\alpha$ | $\beta$ | Situation | Gray | | Diff | | sDiff | |
|---|---|---|---|---|---|---|---|---|
| | | | n | Power | n | Power | n | Power |
| 0.05 | 0.8 | B | 102 | 0.804 | 108 | 0.852 | 116 | 0.790 |
| | | C | 606 | 0.767 | 208 | 0.862 | 220 | 0.886 |
| 0.05 | 0.9 | B | 136 | 0.894 | 144 | 0.925 | 152 | 0.898 |
| | | C | 810 | 0.882 | 278 | 0.919 | 294 | 0.965 |
| 0.01 | 0.8 | B | 152 | 0.793 | 160 | 0.854 | 168 | 0.776 |
| | | C | 900 | 0.776 | 308 | 0.846 | 322 | 0.841 |
| 0.01 | 0.9 | B | 194 | 0.895 | 204 | 0.926 | 212 | 0.889 |
| | | C | 1146 | 0.893 | 392 | 0.931 | 408 | 0.940 |

**Table 5** *Statistical results of the above tests for the two examples*

| Index | Example1 ($\tau$ =41.8 months) | | | | Example2 ($\tau$ =15.3 years) | | | |
|---|---|---|---|---|---|---|---|---|
| | No (95%CI) | Yes (95%CI) | Difference/Ratio[§] (95%CI) | P (statistic) | No radiotherapy (95%CI) | Radiotherapy (95%CI) | Difference/Ratio[§] (95%CI) | P (statistics) |
| CHR | | | 0.86 (0.59,1.25) | 0.053(1.93)[a] | | | 1.14 (0.78,1.65) | 0.503(0.67)[a] |
| SHR | | | 0.60 (0.36,1.00) | 0.049(3.89)[b] | | | 1.45 (0.98,2.14) | 0.072(3.24)[b] |
| RMSTc | 19.72 (18.40,21.03) | 21.55 (14.82,28.28) | 1.83 (-5.03,8.69) | 0.601(0.52)[c] | 8.52 (8.46,8.57) | 9.09 (7.39,10.79) | 0.66 (-1.13,2.28) | 0.510(0.66)[c] |
| RMTL | 15.49 (13.53,17.45) | 9.57 (5.18,13.96) | -5.92 (-10.72, 1.11) | 0.016(2.41)[d] 0.004(3.06)[e] | 2.96 (2.69,3.24) | 4.68 (3.34,6.03) | 1.72 (0.35,3.09) | 0.014(2.46)[d] 0.005(3.01)[e] |

[§]: CHR and SHR are related to CSH and SDH ratio, respectively; RMSTc and RMTL are related to RMSTc difference and RMTL difference, respectively.
[a]: log-rank; [b]: Gray; [c]: RMSTc; [d]: Diff; [e]: sDiff.

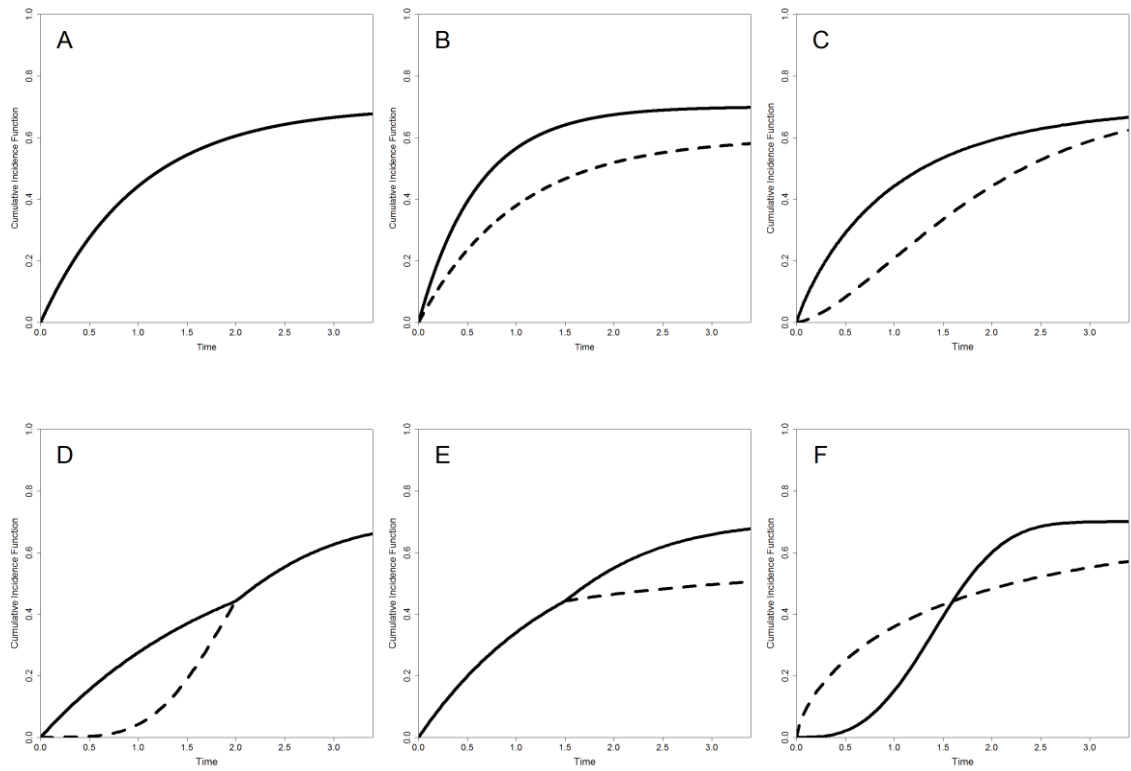

**Figure 1.** Six scenarios in the simulation study — CIF curves for the event of interest.

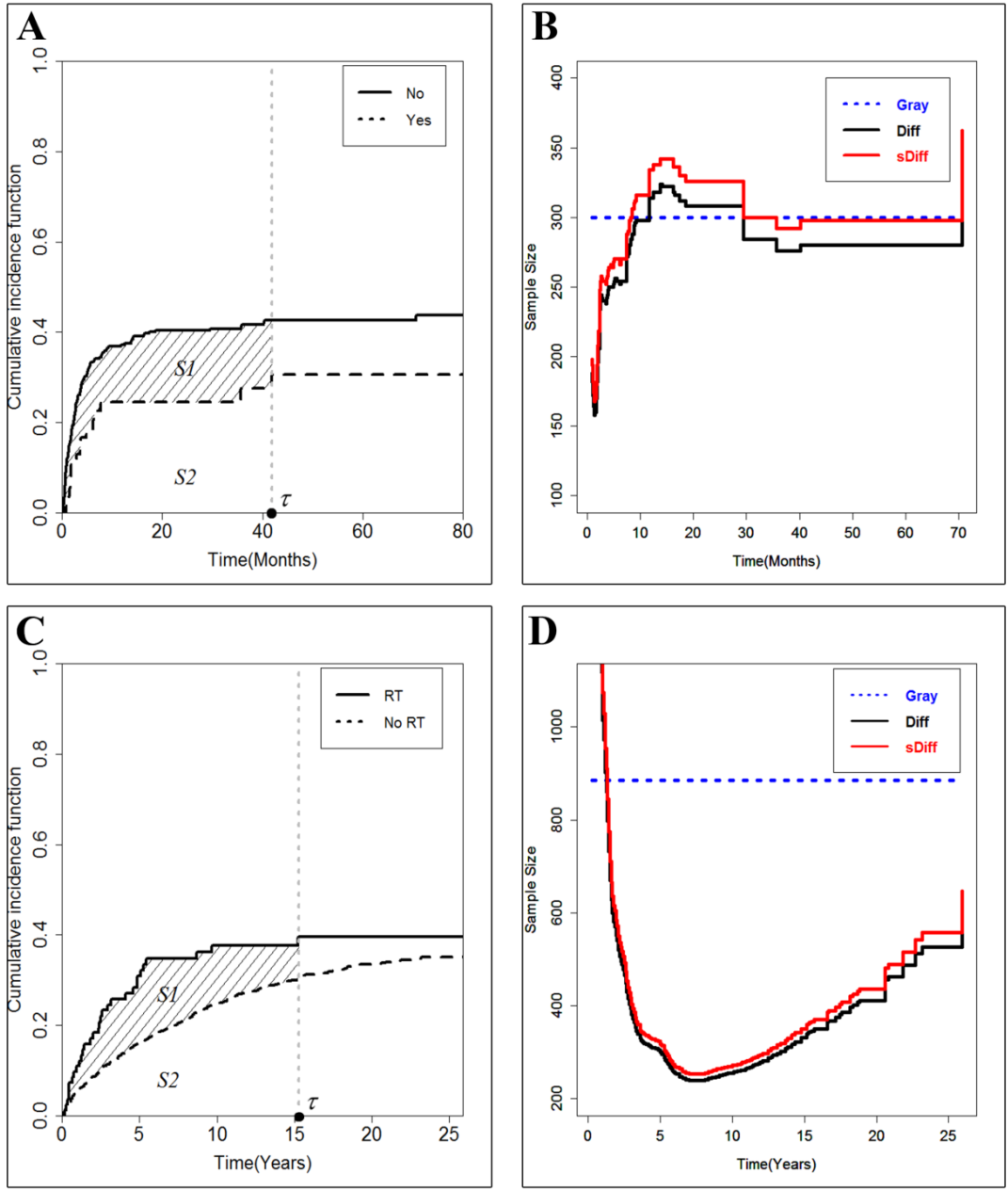

**Figure 2.** CIF curves and calculated sample size for the two examples. A) displays the CIF curves of death from treatment-related causes in example 1. B) displays the calculated sample size change with $\tau$ in example 1. C) displays the CIF curves for death from LL in example 2. D) displays the calculated sample size change with $\tau$ in example 2.